\documentclass[aps,prb,twocolumn,showpacs]{revtex4}

\usepackage{amsmath} \usepackage{epsfig} \usepackage{graphics}

\def\ba{\begin{array}} \def\ea{\end{array}}

\begin{document}


\title{Nonlinear cotunneling through an artificial molecule}


\author{Udo Hartmann} \email[Correspondence to
\\]{hartmann@theorie.physik.uni-muenchen.de}
\author{Frank K. Wilhelm} \affiliation{Sektion Physik and CeNS,
Ludwig-Maximilians-Universit\"at,  Theresienstr. 37, D-80333
M\"unchen, Germany}


\date{\today}

\begin{abstract}
We study electron transport through a  system of two lateral quantum
dots coupled in series.  We consider the case of weak coupling to the leads
and a bias point in the Coulomb blockade. After a generalized
Schrieffer-Wolff transformation,  cotunneling through this system is
described using  methods from lowest-order perturbation theory.  We
study the system for arbitrary bias voltages below the Coulomb energy.
We observe a rich, non-monotonic behaviour of the stationary current
depending on the internal degrees of freedom.  In particular, it turns
out that at fixed transport voltage,  the current through the system
is largest at weak-to-intermediate inter-dot coupling.
\end{abstract}


\maketitle

Quantum dots are prototype systems for studying the properties of
discrete levels embedded in a solid-state environment. Single dots
(``artificial atoms'' \cite{Ashoori}) can be coupled through quantum
point contacts, leading to ``artificial molecules''. Indeed it has
been shown experimentally \cite{Waugh,Blick1,AlexScience}, 
that the eigenstates of
double dot systems are coherent molecular superpositions of single dot
(atomic) states. Unlike real molecules, these dots are readily
contacted and tunable in situ, making them a natural test-bed
for molecular  transport. Double dots have also been proposed as
charge quantum  bits \cite{Blick2,UdoFrank}.

\begin{figure}
\centering \includegraphics[width=0.6\columnwidth]{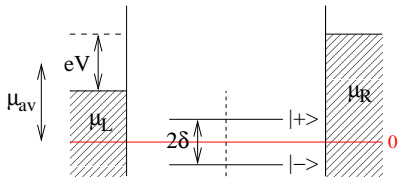}
\caption{Sketch of the considered artificial molecule, where $2\delta$
is the  level splitting and $V$ the bias voltage. The coupling to the
outside leads (hatched areas) is assumed to be small whereas the
inter-dot coupling (dotted  line) can be strong.}
\label{fig:molecule}
\end{figure}

This raises the question, which information on the energy
spectrum and the wavefunctions of the dot can be probed by
transport measurements.  This is only possible if artifacts
induced by the coupling to the leads can be sorted out and the 
double-dot is disturbed as little as possible. 
This is the case, when the coupling to the outside leads is weak  (see
Fig.~\ref{fig:molecule}) and the gates are tuned to the Coulomb
blockade regime \cite{Leo,Wilfried}. In that regime, only states with
a fixed number of electrons are energetically permissible and hence
sequential tunneling is suppressed. The leading transport 
mechanism in this case is cotunneling \cite{Nazarov}, the coherent transfer 
of two electrons via virtual levels in the dots. Our work stands
between studies focusing on sequential tunneling \cite{Stoof} and 
work on {\em linear response} in the Kondo regime\cite{Laszlo}. 
The properties of cotunneling currents as a spectroscopic tool for the 
spectrum of quantum dot system
have recently been studied in exquisitely  controlled experiments on
systems similar to ours \cite{Silvano,AlexScience}.

In this paper, we analyze a serial  configuration of lateral quantum
dots in the cotunneling regime. We study finite voltages up to the order
of the charging energy i.e.\ do not restrict ourselves to linear response. 
We  find a rich nonmonotonic structure in the
current as a function of  the dot parameters. In particular, we find a
pronounced crossover indicating the opening of an inelastic transport
channel, which leads to the surprising result, that a too strong
interdot coupling actually inhibits charge transport. We analyze the
influence of the asymmetry of the dots on the current.

In the Coulomb blockade regime \cite{Leo,Wilfried}, the relevant
Hilbert space is spanned by four basis states $|i,j\rangle$,
$i,j\in\lbrace0,1\rbrace$,  which denotes $i$ and $j$  additional
electrons (as compared to an appropriate neutral state)  on the left
and right dot respectively.  We study the situation where the gate
voltages of the single dots are very close to  each other and the
inter-dot coupling is, although appreciable, much smaller than the
single dot addition energy. Thus, the subspace spanned by the  two
states $|1,0\rangle$ and $|0,1\rangle$ is energetically most
favorable. The next closest states  $|v_0\rangle=|0,0\rangle$ and
$|v_2\rangle=|1,1\rangle$ are  outside the transport window and serve
as virtual states \cite{Nazarov}.  States with higher dipolar moment
are energetically even less favorable  due to the high charging energy
of the individual dots.

The Hamiltonian of this system can be written as
\begin{eqnarray}
H & = & H_{0} + H_{1}\\ H_{0} & = & H_{{\rm sys}} + H_{{\rm res}} \\
H_{{\rm sys}} & = & \varepsilon_{\rm as} (\hat{n}_l-\hat{n}_r) -
\varepsilon_\alpha \hat{n}_{v_0}+\varepsilon_\beta \hat{n}_{v_2} \nonumber
\\ && + \gamma \sum_{n}
(a^{L\dagger}_{n}a^{R}_{n}+a^{R\dagger}_{n}a^{L}_{n}) \\ H_{{\rm res}}
& = & \sum_{\vec{k}}
\varepsilon^{L}_{\vec{k}}b^{L\dagger}_{\vec{k}}b^{L}_{\vec{k}} +
\sum_{\vec{k\prime}}
\varepsilon^{R}_{\vec{k\prime}}b^{R\dagger}_{\vec{k\prime}}b^{R}_{\vec{k\prime}}
\\ H_{1} & = & t_{c} \sum_{\vec{k},n}
(a^{L\dagger}_{n}b^{L}_{\vec{k}}+a^{L}_{n}b^{L\dagger}_{\vec{k}})
\nonumber \\  && + t_{c} \sum_{\vec{k\prime},n}
(a^{R\dagger}_{n}b^{R}_{\vec{k\prime}}+a^{R}_{n}b^{R\dagger}_{\vec{k\prime}})
\ .
\end{eqnarray}
Note, that the sum over dot states $n$ only runs over the restricted
Hilbert space described above.  $H_{0}$ describes the isolated {\em
double-dot}  ($H_{{\rm sys}}$) and the leads ($H_{{\rm res}}$),
whereas the tunneling part $H_{1}$ describes the coupling of each dot
to its lead and will be treated as a perturbation.  $\hat{n}_{l/r}$
are the number operators counting additional electrons on either
dot. The asymmetry energy $\varepsilon_{\rm
as}=(\varepsilon_{l}-\varepsilon_{r})/2$ describes half of the difference
between the energy level for the additional electron  in left dot
($\varepsilon_{l}$) and the corresponding energy level in the right dot
($\varepsilon_{r}$), which can be tuned through the gate voltages.
$\varepsilon_{\beta}$ and $\varepsilon_{\alpha}$ are the charging energies
towards the higher level $|v_2\rangle$ and  the lower level
$|v_0\rangle$ respectively.  $\gamma$ is the tunable inter-dot
coupling strength.  The $a^{(\dagger)}$s and $b^{(\dagger)}$s denote
electron creation/annihilation operators in the dots and leads.  In
$H_{1}$, the symbol $t_{c}$ represents the tunnel matrix element
between the dots and the leads. It is independent of the energies in
the  double-dot system and the corresponding sequential tunneling rate
$\hbar\Gamma=2\pi t_c^2 N(\varepsilon_{\rm F})$ 
should be  small compared to the
internal energies. $N(\varepsilon_{\rm F})$ 
is the density of states in the leads taken at the Fermi energy.
We restrict our analysis to spin-polarized electrons, these can be
polarized by an appropriate in-plane magnetic field.
Fig.~\ref{fig:molecule} shows a sketch of the system.
In Fig.~\ref{fig:molecule}, $V=\mu_{R}-\mu_{L}$ is the bias voltage
between the two leads (hatched) and $2\delta=2\sqrt{\varepsilon_{\rm
as}^2+\gamma^2}$ is the level splitting in the  molecular two-state
system (TSS).

Pursuing our aforementioned objective, we take the inter-dot coupling
$\gamma$ into account to all orders by diagonalizing $H_{\rm sys}$ and
transforming $H_{\rm 1}$ into the new basis. Already now,
there is no simple selection rule or symmetry of the coupling of the
states to the leads  any more. We want to use  well-established
tools of lowest order perturbation theory for both finding the density
matrix of the system and evaluating the current.  In order to
capture cotunneling by this approach,  we perform a  Schrieffer-Wolff
transformation \cite{Schrieffer} up to second order,  i.e.\ we take
into account all indirect transitions between arbitrary final  and
initial states of the dot which involve only a {\em single} intermediate
state. This takes the transformed Hamiltonian
into the  generic form
\begin{eqnarray}
\tilde{H}_{I} & = & \sum_{c,d}\limits \alpha_{c}^{\dagger}\alpha_{d} \Big[
\sum_{Y,Y',\vec{k},\vec{k'}}\limits H^{Y,Y'}_{\vec{k},\vec{k'},c,d}
b^{Y\dagger}_{\vec{k}} b^{Y'}_{\vec{k'}} + \nonumber \\
&& + \sum_{Y,Y',\vec{k},\vec{k'}}\limits
H^{Y,Y'}_{\vec{k},\vec{k'},c,d} b^{Y}_{\vec{k}} b^{Y'\dagger}_{\vec{k'}}\Big] ,
\label{eq:transhamil}
\end{eqnarray}
where the $H^{Y,Y'}_{\vec{k},\vec{k'},c,d}$ are Schrieffer-Wolff
amplitudes and  $c,d=\pm$ denote the two molecular levels,
$\alpha^{(\dagger)}_{c/d}$ the associated molecular operators and $Y,Y'$
the position of the electrons involved  in these processes. Due to the
molecular nature of the double dot eigenstates, all the amplitudes are
finite and composed out of a huge number of contributions with no
particular  symmetry. The perturbation theory formula for this general
case can  be found e.g.\ in Ref.\ \onlinecite{CohenTannoudij} and is
worked out in more detail in Refs.\ \onlinecite{UdoFrank} and 
\onlinecite{Diplomarbeit}. In eq.\
(\ref{eq:transhamil}), we  have taken matrix elements in the
double-dot eigenbasis only whereas we stick to second-quantized
notation in the leads, because this notation readily connects to the
formalism used later on.

The stationary density matrix is found using the well-established and
controlled Bloch-Redfield theory \cite{Agyres}. This is a systematic
technique for deriving generalized master equations within Born
approximation in $\tilde{H}_I$,  eq.\ (\ref{eq:transhamil}), which
includes all relevant  non-markovian parts. This approach has been
shown \cite{Hartmann} to be numerically equivalent to formally exact
path integral methods for the Spin-Boson model in the weak-coupling limit.
The Redfield equations \cite{Weiss} for the
elements of the reduced density  matrix $\rho$ in the molecular basis
read
\begin{equation}
\dot{\rho}_{nm}(t)=-i\omega_{nm}(t)\rho_{nm}(t)-\sum_{k,l}\limits
R_{nmkl}\rho_{kl}(t) \ ,
\label{redfield}
\end{equation}
where $\omega_{nm}=(E_n-E_m)/\hbar$ are the appropriate energy
splittings  and $R_{nmkl}$ are the elements of the  Redfield
tensor. They are composed out of golden rule rates involving
$\tilde{H}_I$ from eq.\ (\ref{eq:transhamil}). $n$, $m$, $k$ and $l$
can be  either $+$ (molecular excited state) or $-$ (molecular ground
state). The $E$s are the eigenenergies of the two molecular
states. Due to the lack of symmetry,  this leads to a huge number of
processes contributing to each  term \cite{Diplomarbeit}.
We are only interested in stationary solutions here. A full treatment
of the simple case with $\gamma=0$ can be found in Ref.\
\onlinecite{UdoFrank}.

The current is derived from the standard formula \cite{Mahan}
\begin{equation}
I(t)=-e \frac{i}{\hbar} \int_{-\infty}^{t}\limits dt'
\langle[\dot{N}_L(t),\tilde{H}_I(t')]\rangle \ ,\label{eq:current}
\end{equation}
where $N_L$ is the particle number operator on the left dot in the
interaction representation and the transformed interaction
Hamiltonian $\tilde{H}_I$ from eq.\ (\ref{eq:transhamil}) is also
taken in the interaction picture. Carrying out the integration in equation
(\ref{eq:current}) and rotating back to the Schr\"odinger picture, we get a
time-independent expression for the current $I$. Using the stationary 
occupation probability of the molecular ground ($\rho_{\rm --,st}$) or excited
state ($\rho_{\rm ++,st}$), we obtain for the expectation value of the
stationary current
\begin{equation}
I_{\rm st} ={\rm tr} (\rho_{\rm st}I)  =\rho_{\rm ++,st}I_{++} +
\rho_{\rm --,st}I_{--} \ ,
\label{eq:st_current}
\end{equation}
where we find from balancing relaxation processes in the
Bloch-Redfield equation, eq.\ (\ref{redfield})
\begin{equation}
\rho_{\rm ++,st} =  \frac{R_{++--}}{R_{++--}-R_{++++}} \quad \rho_{\rm
--,st} = \frac{R_{--++}}{R_{--++}-R_{----}} \ .
\end{equation}

\begin{figure}[h]
\begin{center}
\parbox{2.5cm}{ (a) \\ \includegraphics[width=2.5truecm]{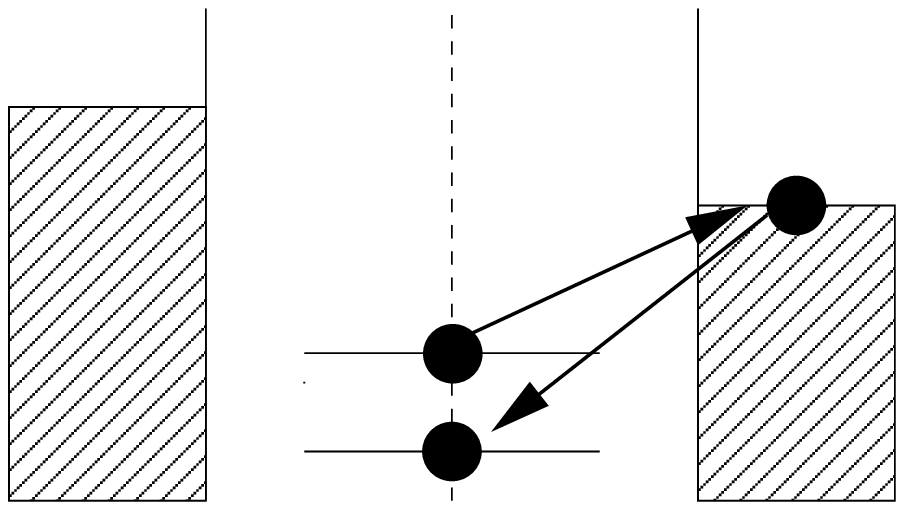}}
\hfill
\parbox{2.5cm}{ (b) \\
\includegraphics[width=2.5truecm]{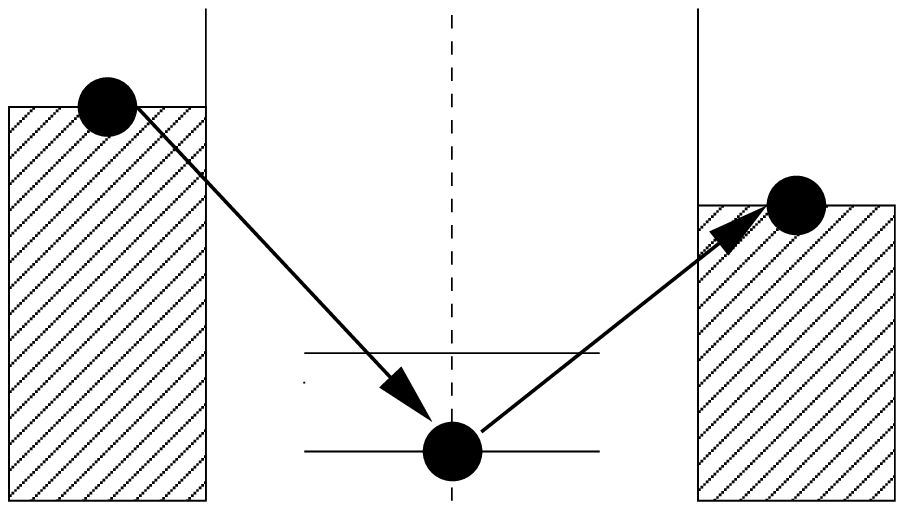}}  \hfill
\parbox{2.5cm}{ (c) \\ \includegraphics[width=2.5truecm]{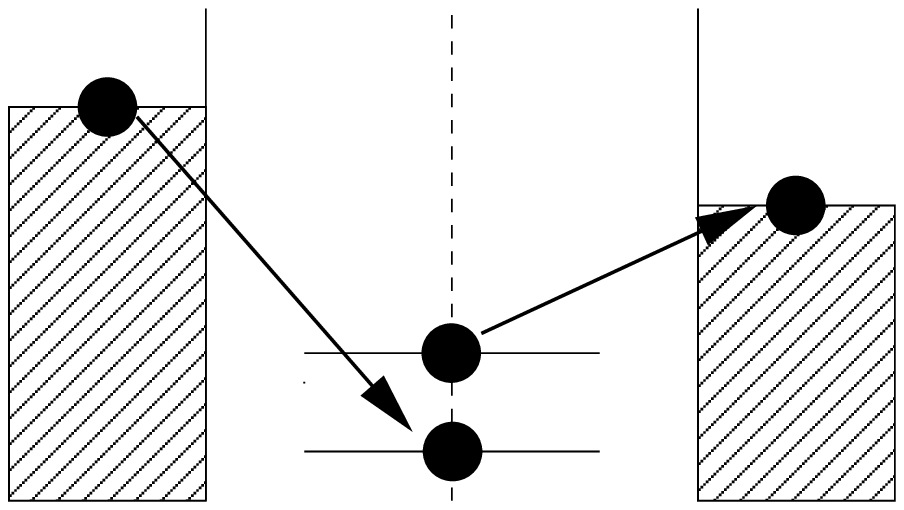}}
\end{center}
\caption{Examples for relevant processes: (a) a relaxation process
without  current, (b) current without relaxation (only dephasing) and
(c) a process that carries current and also relaxes the system.}
\label{procs}
\end{figure}

The current amplitudes $I_{++}$ and $I_{--}$ in eq.\ (\ref{eq:st_current}) are
of the same
form as the contributions to the Redfield tensor. We 
emphasize, that the choice of processes from all
possibilities is very distinct. As an example,  Fig.~\ref{procs}
displays a variety of possible processes in such a double-dot
system. Processes of the type displayed  in Fig.~\ref{procs}  (a)
contribute to the relaxation but do not carry current, (b) shows a
process which carries current but does not relax the state, and (c)
relaxes {\em and} carries current.  The phase information of the
quantum state is lost in all three pictures of
Fig.~\ref{procs}. Consequently, one must not confuse cotunneling rates with
relaxation rates.

We now turn to the discussion of the results. All internal energies
$\varepsilon_{\rm as}$ and $\gamma$ are normalized in units of the bias
voltage  $V$,  the stationary current $I_{\rm st}$ in terms of
$I_{0}=e\Gamma$.
\begin{figure}[h]
\centering \includegraphics[width=0.7\columnwidth]{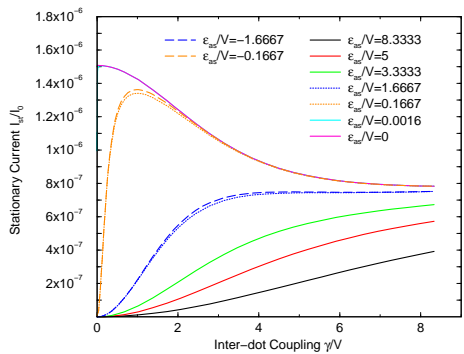}
\caption{Stationary current $I_{\rm st}/I_{0}$ for different 
$\varepsilon_{\rm as}/V$ as a function of the coupling  
$\gamma/V$.
(with  $T=140$~mK, $V=5.170$~$\mu$V and $\mu_{\rm
av}=(\mu_{R}+\mu_{L})/2=75.832$~$\mu$eV and $\Gamma=1$~GHz.).}
\label{fig:cur_g}
\end{figure}

In Fig.~\ref{fig:cur_g} the current at fixed bias voltage as a
function of the inter-dot coupling is shown.
The sign of $\varepsilon_{\rm as}$ plays a role, as one can see above,
for an intermediate $\varepsilon_{\rm as}$-regime. This effect is more
pronounced in $I(V)$, see Ref.\  \cite{Diplomarbeit}.
Close to $\gamma=0$, the curves all turn to zero, because at that
point the dots are disconnected and no current can flow.  However, a
number of curves, the ones with $\varepsilon_{\rm as}/V<1$, exhibit an
intermediate maximum at low $\gamma$ next to a very sharp minimum at
$\gamma=0$,  which sometimes is hardly resolved.  At high
$\gamma\gtrsim V$,  the stationary current saturates into a value,
which for our parameters turns out to be about $I_{\rm 0,st}/I_{0}=7.5
\cdot 10^{-7}$. Remarkably, this is half the value of the current at
the aforementioned  low-$\gamma$-Maximum. This is the central result
of this paper.

These regimes can be classified in terms of the level splitting
$2\delta$\cite{Wegewijs}: At $V<2\delta $, the energy $V$ supplied from the 
leads is
only sufficient to use one of the molecular states for transport 
(elastic cotunneling) whereas at $V>2\delta$,
both states participate and also inelastic processes contribute, i.e.\
there is a second current channel, which carries the same contribution
of $I_{\rm 0}$.  The crossover naturally occurs at  $\gamma =
\sqrt{\frac{V^2}{4}-\varepsilon_{\rm as}^2}$, which can only be reached
if $\varepsilon_{\rm as}/V<1/2$.  As long as $\gamma$ is not too low,
the coupling to the leads is the limiting element for the current
flow. Only  if $\gamma < \varepsilon_{\rm as}$, the double-dot eigenstates become
localized and the inter-dot coupling becomes the current bottleneck. 
Consequently, associated dips have a half-width of
$\varepsilon_{\rm as}$ for low temperatures and bias voltages and can
thus be extremely narrow. We would like to remark that the notion of
transport ``channels'' is appropriate here, because cotunneling
is a coherent transport process. 

\begin{figure}[h]
\centering \includegraphics[width=0.7\columnwidth]{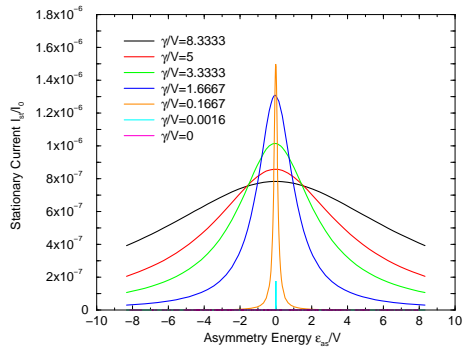}
\caption{Stationary current $I_{\rm st}/I_{0}$ for different values of
$\gamma/V$ as a function of the asymmetry energy $\varepsilon_{\rm as}/V$ (with
$T=140$~mK, $V=5.170$~$\mu$V and $\mu_{\rm av}=75.832$~$\mu$eV and 
$\Gamma=1$~GHz).}
\label{fig:cur_e}
\end{figure}
Fig.~\ref{fig:cur_e} shows the dependence of the stationary current on
$\varepsilon_{\rm as}/V$. It confirms the
interpretation of Fig.~\ref{fig:cur_g}.  The plot
is only weakly asymmetric to $\varepsilon_{\rm as}/V=0$.  At zero asymmetry,
$\varepsilon_{\rm as}/V=0$,  the condition for charge transport is ideal,
$\sqrt{\frac{V^2}{4}-\varepsilon_{\rm as}^2}$ has its maximum and
therefore the current is only governed by the inter-dot coupling
$\gamma/V$, resulting in a zero-asymmetry maximum.

Still, all three transport regimes can be recognized in
Fig.~\ref{fig:cur_e}.  The $\gamma/V=0$ curve shows,  that the
stationary current $I_{\rm st}/I_{0}$ is exactly zero as expected.
For growing, but small values of $\gamma/V$, the maximum at
$\varepsilon_{\rm as}/V=0$ reaches the highest value  $I_{\rm
st}/I_{0}=2I_{\rm 0,st}/I_{0}$ at about $1.5 \cdot 10^{-6}$ (like in
Fig.~\ref{fig:cur_g}), corresponding to two open transport channels
(elastic {\em and} inelastic). If we raise $\gamma/V$ further, the
height of the peak goes down again and saturates at $I_{\rm
st}/I_{0}=I_{\rm 0,st}/I_{0} \approx 7.5 \cdot 10^{-7}$, corresponding
to only the elastic channel being open.
\begin{figure}[t]
\centering \includegraphics[width=6truecm]{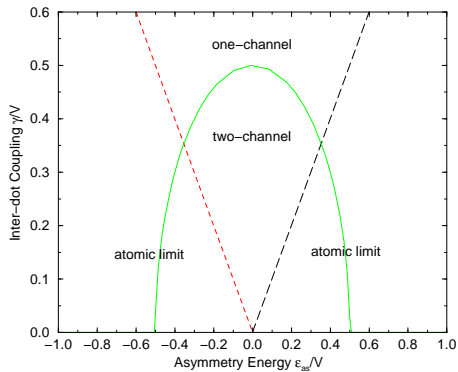}
\caption{Limits for the three transport regimes with $V=5.170$~$\mu$V}
\label{fig:phase}
\end{figure}

The three transport regimes are summarized in Fig.~\ref{fig:phase}: i)
the atomic limit (no transport) $\gamma < \varepsilon_{\rm as}$,  ii) the
two-channel case (inelastic cotunneling)  $\varepsilon_{\rm as} < \gamma
< \sqrt{\frac{V^2}{4}-\varepsilon_{\rm as}^2}$; and iii) the one-channel
case (elastic cotunneling)  $\gamma >
\sqrt{\frac{V^2}{4}-\varepsilon_{\rm as}^2}$.

These conditions show that indeed cotunneling can be  used as a tool
for investigating the energy spectrum of an undisturbed artificial
molecule \cite{AlexScience}.  The cross-over between the elastic and
the inelastic cotunneling in dependence of the applied bias voltage
has recently been observed \cite{Silvano}.  A similar conclusion was
found in Ref.\ \onlinecite{Wegewijs}.

Although the notion of (elastic and inelastic) cotunneling was already
introduced very early\cite{Nazarov}, its consequences for realistic 
quantum dot systems have only been discussed very recently\cite{Wegewijs}, 
along with
detailed and accurate
experiments on small semiconductor quantum dots 
\cite{Silvano,AlexScience} becoming available. The sharp crossover between
elastic and inelastic cotunneling, which we discuss, has been identified
in a vertical quantum dot\cite{Silvano} by changing the transport
voltage.
Ref.\ \onlinecite{AlexScience} studies cotunneling in  a parallel
double-dot topology, using again cotunneling and the
elastic-to-inelastic crossover as a spectroscopic tool and tuning the
inter-dot coupling {\em in  situ}. In both cases, the narrow regime of
decoupled dots would not have been accessible through a conductance
measurement. Some of the experimental issues have been theoretically
addressed in Ref.\ \onlinecite{Wegewijs}. In that case, however, the
behavior of a single multilevel dot system was modelled with
phenomenological couplings to  the leads, whereas we take a realistic
model and only by this manage to predict effects which e.g.\ depend on
the serial dot topology of the sample.  Note, that parts of the
double-dot literature focus on phonon/photon   assisted transport (see
e.g.\ Refs.\ \onlinecite{Qin,Oosterkamp} for  experiments  and Refs.\
\onlinecite{Hazelzet,Brandes1} for theory). Unlike  Ref.\
\onlinecite{Juergen}, we concentrate on the Coulomb blockade regime
and do not consider cotunneling at resonance. In Ref.\
\onlinecite{Loss}, a different approach to the problem  was developed,
in which the master equation is carried to second order instead of
using a Schrieffer-Wolff transformation, and a few setups simpler  than
ours are studied.  Our approach does not require the molecule to be
artificial, in principle it can be applied to ``real'' molecules
\cite{Weber}. In contrast to the approach in Ref.\
\onlinecite{Cuevas}, it permits to take into account charging effects, 
however, the Schrieffer-Wolff transformation is clearly a
laborious step for larger systems.

To conclude, we analyzed the stationary coherent cotunneling current 
$I_{\rm st}$
through a
double quantum dot system or artificial molecule.  
As a function of the inter-dot coupling strength it displays
a rich, non-monotonic structure, which enables us to perform
``molecular cotunneling spectroscopy''. Strikingly, we have shown that
at fixed bias voltage, the current is highest, if the dots are
weakly-to-intermediately  connected, such that the inter-dot coupling
is at  least as strong as the coupling to the leads, but the splitting
of the  molecular wavefunctions is still smaller than the transport
voltage.

We thank J.\ von Delft, J.\ K\"onig, L.\ Borda, M.\ Sindel, A.W.\
Holleitner, A.K.\ H\"uttel and E.M.\ H\"ohberger for clarifying
discussions. We acknowledge financial support from ARO,
Contract-No. P-43385-PH-QC.

\end{document}